\documentclass[seceq]{ptptex}

\usepackage{graphicx}
\usepackage{epstopdf}
\usepackage{pstricks}

\newcommand{\be}{\begin{equation}}
\newcommand{\ee}{\end{equation}}
\newcommand{\ba}{\begin{eqnarray}}
\newcommand{\ea}{\end{eqnarray}}
\newcommand{\ex}{{\rm e}}
\newcommand{\nn}{\nonumber}

\newcommand{\eq}{Eq.~}

\newcommand{\fig}{Fig.~}

\title{
Strong coupling series for QCD at finite temperature and density
}

\author{
Jens \textsc{Langelage}$^1$ and Owe \textsc{Philipsen}$^2$%
}

\inst{
$^1$ Fakult\"at f\"ur Physik, Universit\"at Bielefeld, \\
33615 Bielefeld, Germany\\
$^2$ Institut f\"ur Theoretische Physik, 
Goethe-Universit\"at Frankfurt\\  
Max-von Laue-Str. 1, 60438 Frankfurt am Main, Germany
}

\abst{
We discuss the use of strong coupling expansions for
Yang-Mills theory and QCD at finite temperature and density.
In particular we consider the onset of temperature effects for
the free energy and screening masses, derive the hadron 
resonance gas model 
from first principles and compute the
weakening of the deconfinement transition with chemical potential.
}

\begin{document}

\maketitle

\section{Introduction}
Strong coupling expansions 
were among the first tools to study 
lattice gauge theory and produced
analytical insight into confinement and glueball masses. Here we 
discuss some
recent extensions of these techniques 
to finite temperature and
density \cite{sc1}.
Strong coupling expansions are complementary to those in weak coupling
and yield convergent series with a finite radius of convergence.
They are the only analytic tool to address the confined phase of QCD from
first principles,
allowing to study the onset of finite temperature effects at
low temperatures. Furthermore, with strong coupling series of sufficient length we may hope 
to establish a connection between QCD and models resulting in the strong coupling limit, which
are often used for studies of nuclear matter at finite baryon density \cite{scl1,scl2}.
 
\section{Equation of state and screening masses for Yang-Mills}

Starting point is the YM partition function 
using the Wilson action
\be
Z=\int \,DU\,\exp -S_{YM}[U],\quad S_{YM}=\sum_p\frac{\beta}{2N}\left( \mathrm{Tr}\, U+\mathrm{Tr}\,U^{\dagger}-2N\right),
\ee
with the lattice gauge coupling $\beta=\frac{2N}{g^2}$.
An expansion of $\tilde{f}\equiv-\frac{1}{\Omega}\ln Z$ 
in $\beta$ proceeds by expanding in 
group characters $\chi_r(U)$ followed by a
cluster expansion,
\cite{Montvay:1994cy} 
\be
\tilde{f}=-6\ln\,c_0(\beta)-\frac{1}{\Omega}
\sum_{C=(X_i^{n_i})}\,a(C)\prod_i\Phi(X_i)^{n_i},\quad
\Phi(X_i)=\int DU\prod_{p\in X_i} d_{r_p} a_{r_p} \chi_{r_p}(U),
\label{free}
\ee
where $\Omega=V\cdot N_t$ is the lattice volume, $d_r$ and $a_r$
dimension and expansion coefficient of representation $r$, and $c_0$ is 
the expansion coefficient of the trivial representation. 
The standard expansion parameter to express results in is 
conventually the coefficient of the fundamental representation,
$u\equiv a_f$.
The combinatorial factor $a(C)$  
equals $1$ for clusters $C$ which consist of only one 
so-called polymer $X_i$. 
The (vacuum) quantity in \eq(\ref{free}) is customarily 
called a free energy density, because
the path integral corresponds to a partition function if one 
formally identifies $\beta$ with
$1/T$. Here we are interested in a physical temperature $T=1/(aN_t)$, 
realised by compactifying the
temporal extension of the lattice. 
The physical free energy is then obtained by subtracting  
the divergent vacuum free energy,
\begin{equation}
f(N_t,u)=\tilde{f}(N_t,u)-\tilde{f}(\infty,u),
\end{equation}
and the pressure is $P=-f$.
Group integrals are evaluated using the formulae
\begin{equation}
\int dU \chi_r(UV)\chi_r(WU^{-1})=\chi_r(VW),\quad \int dU \chi_r(U)=\delta_{r,0}.
\end{equation}
Due to the latter the contributing graphs $X_i$ have to be objects with a closed surface. 

\begin{figure}[t]
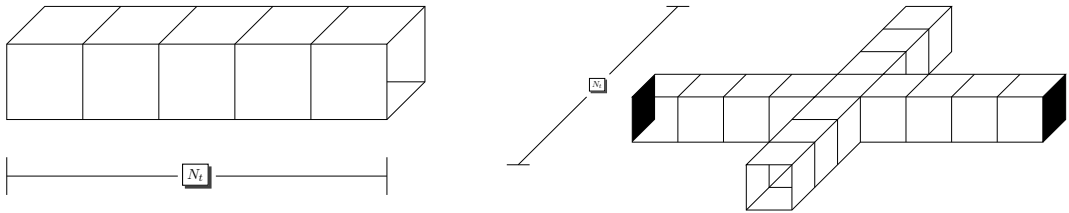

\vspace*{2.3cm}
\begin{minipage}{7.5cm}
\hspace{3cm}
\scalebox{0.25}{
\pspolygon[linecolor=black](-10,7)(-10,3)(10,3)(12,5)(12,9)(-8,9)
\psline(-10,7)(10,7)(10,3)
\psline(10,7)(12,9)
\psline(-6,3)(-6,7)(-4,9)
\psline(-2,3)(-2,7)(0,9)
\psline(2,3)(2,7)(4,9)
\psline(6,3)(6,7)(8,9)
\psline(12,5)(10,5)
\psline(-10,0)(-1,0)\psline(-10,1)(-10,-1)
\psline(1,0)(10,0)\psline(10,1)(10,-1)\rput(0,0){\scalebox{2}{\psshadowbox{$N_t$}}}}
\end{minipage}
\begin{minipage}{7.5cm}
\hspace{4cm}\scalebox{0.15}{\pspolygon*[fillcolor=red](-18,7)(-18,3)(-16,5)(-16,9)\pspolygon*[fillcolor=red](18,7)(18,3)(20,5)(20,9)\pspolygon[linecolor=black](-16,9)(0,9)(6,15)(10,15)(4,9)(20,9)(18,7)(2,7)(-4,1)(-8,1)(-2,7)(-18,7)\psline(-18,7)(-18,3)(-6,3)\psline(-8,1)(-8,-3)(-4,-3)(2,3)(18,3)(18,7)\psline(18,3)(20,5)(20,9)\psline(-4,-3)(-4,1)\psline(2,3)(2,7)(4,9)\psline(10,15)(10,11)(8,9)\psline(-12,9)(-14,7)(-14,3)\psline(-8,9)(-10,7)(-10,3)\psline(-4,9)(-6,7)(-6,3)\psline(4,9)(0,9)(-2,7)(2,7)\psline(-6,3)(-2,3)(-2,-1)\psline(-4,5)(0,5)(0,1)\psline(2,11)(6,11)(6,9)\psline(4,13)(8,13)(8,9)\psline(6,3)(6,7)(8,9)\psline(10,3)(10,7)(12,9)\psline(14,3)(14,7)(16,9)\psline(-8,-3)(-4,1)\psline(-6,1)(-6,-1)(-4,-1)\psline(-28,1)(-22,7)\psline(-29,1)(-27,1)\psline(-20,9)(-14,15)\psline(-15,15)(-13,15)\rput(-21,8){\scalebox{2}{\psshadowbox{$N_t$}}}}
\end{minipage}
\vspace*{0.3cm}
\caption{Left: LO graph for the free energy. Right: LO graph
for the screening mass.}
\label{fig_tube}
\end{figure}

The graph contributing to the leading order of the free energy 
density is a tube of length $N_t$ with a cross-section of one 
single plaquette, Fig.~\ref{fig_tube}. 
Summing over all such graphs on the lattice, their contribution is 
\be
P^{LO}(N_t,u)=\frac{6}{N_t}u^{4N_t}\;.
\ee
Thus, the strong coupling limit at $\beta=0$ (and thus $T=0$) has zero free enery density or pressure,
as expected. Moreover, we see that the pressure rises very slowly with $T$ as the leading order
starts at a high power in the expansion parameter only. 
Corrections to the free energy density through
order $u^8$ with respect to the leading order term for various $N_t$
have been
calculated for SU(2) \cite{sc1} and for SU(3) \cite{sc3}.  

Next, let us turn to screening masses. The one with the quantum numbers of the lowest lying  glueball   is defined by the spatial correlation 
function of plaquettes,
\begin{equation}
C(z)=\langle\mathrm{Tr}\,U_{p_1}(0)\,\,\mathrm{Tr}\,U_{p_2}(z)
\rangle=N^2\frac{\partial^2}{\partial\beta_1\partial\beta_2}\ln\,
Z(\beta,\beta_1\beta_2)\bigg\vert_{\beta_{1,2}=\beta}.
\end{equation}
At zero temperature the exponential decay is the same as for correlations in the time direction, and thus
determined by the glueball masses. At finite temperature, 
the LO
graph for the difference to the vacuum mass is shown in 
Fig.~\ref{fig_tube} and gives
\be
\Delta m^{LO}(N_t,u)=m(N_t)-m(\infty)
=-\frac{2}{3}N_tu^{4N_t-6}
\ee
The finite T effect is to lower the screening 
mass in the confined phase and  
is suppressed by high orders in the strong coupling. This explains
why in the confined phase it is close to the
vacuum glueball mass as observed by Monte Carlo \cite{dg}.

\section{QCD with heavy quarks}

Wilson fermions can be included by means of an expansion in the hopping 
parameter $\kappa_f=(2am_f+8)^{-1}$.
Defining the usual hopping matrix $M$ \cite{Montvay:1994cy},
Grassmann integration over fermion fields gives to leading
order in $\kappa_f$ per flavour
\begin{eqnarray}
-S_q^f
=\sum_{l}\,\frac{\kappa_f^l}{l}\,\mathrm{tr}\,M^l(\mu)
= -(2\kappa_f)^{N_t}\sum_{\bf x} \left( \ex^{\mu N_t}L({\bf x})
+\ex^{-\mu N_t}L^\dag({\bf x})\right)
+\ldots
\label{fact}
\end{eqnarray}
Kronecker deltas in the hopping matrix $M$ force the sum to 
extend solely over closed loops on the lattice. 
Again the finite temperature effects are in the 
difference between finite and infinite $N_t$, and 
hence in the loops winding through the temporal boundary. 
Thus the leading order heavy 
fermionic contributions to the effective action can be written as
in the last equation, which only holds for finite $T$, and
the dots represent loops that wind more than once.

Now we calculate again the free energy density, 
this time performing character expansions in both $U$ 
and $L$. 
Expanding all terms up to ${\cal{O}}(\kappa^{3N_t})$ and 
doing the group integrals we get for two flavours $u,d$
\begin{eqnarray}
P&=&\frac{1}{N_ta^4}\bigg\lbrace4(2\kappa_u)^{2N_t}+8(2\kappa_u2\kappa_d)^{N_t}+4(2\kappa_d)^{2N_t}\bigg\rbrace\nonumber\\
&+&\frac{1}{N_ta^4}\bigg\lbrace4(2\kappa_u)^{3N_t}+6\big[(2\kappa_u)^22\kappa_d\big]^{N_t}\nonumber\\
&+&6\big[2\kappa_u(2\kappa_d)^2\big]^{N_t}+4(2\kappa_d)^{3N_t}\bigg\rbrace\Big(\mathrm{e}^{3a\mu N_t}+\mathrm{e}^{-3a\mu N_t}\Big)
\end{eqnarray}
If we now recognise the hadron masses to leading order hopping
expansion as 
\ba
\mbox{Mesons:}\qquad\quad am_{f\bar{f^\prime}}&=&-\ln2\kappa_f-\ln2\kappa_{f^\prime}\\
\mbox{Baryons:}\qquad am_{ff^\prime f^{\prime\prime}}&=&-\ln2\kappa_f-\ln2\kappa_{f^\prime}-\ln2\kappa_{f^{\prime\prime}},
\ea 
we are able to rewrite this as
\begin{eqnarray}
P&=&\frac{1}{N_ta^4}\left\lbrace\sum_{0^-}\mathrm{e}^{-m\left(0^-\right)N_t}+3\sum_{1^-}\mathrm{e}^{-m\left(1^-\right)N_t}\right\rbrace\nonumber\\
&+&\frac{1}{N_ta^4}\Bigg\lbrace4\sum_{\frac{1}{2}^+}\mathrm{e}^{-m\left(\frac{1}{2}^+\right)N_t}+8\sum_{\frac{3}{2}^+}
\mathrm{e}^{-m\left(\frac{3}{2}^+\right)N_t}
\Bigg\rbrace\cosh\big(\mu_BN_t\big),\label{eq_hrg}
\end{eqnarray}
which is nothing but the pressure of a hadron resonance gas.
We have thus derived from first principles that the latter
arises as an effective theory for QCD in the strong coupling regime.
Note that this is a generic feature holding also for Yang-Mills
theory, which can be represented as a glueball gas \cite{sc1,sc3}.

\section{The deconfinement phase transition}

\begin{figure}[t]
\begin{minipage}{0.5\textwidth}
\begin{eqnarray}
N_f=1:\qquad m_c/T&=&2.08(7),\nn\\
N_f=2:\qquad m_c/T&=&2.78(7),\nn\\
N_f=3:\qquad m_c/T&=&3.17(10).\nn
\end{eqnarray}
\end{minipage}
\begin{minipage}{0.5\textwidth}
\includegraphics[width=0.7\textwidth]{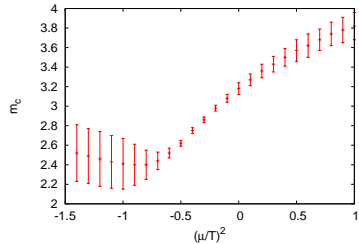}
\end{minipage}
\caption[]{The critical quark mass $m_c(\mu^2)$ for $N_t=1$. 
Left: $\mu=0$. 
Right: $N_f=3$. 
Error bars represent the scatter for different Pad\'e-approximants.} 
\label{fig_hc_mu}
\end{figure}

It is well known that the deconfinement transition is first order 
in the pure
gauge limit, $\kappa=0$, and weakens for finite quark masses until it disappears at some 
critical $\kappa_c$, which represents an upper bound on 
the radius of convergence for our strong coupling series. 
One may extract this quantity
from the Polyakov
loop susceptibility, 
\be
\chi_L(J)=\frac{1}{V}\frac{\partial^2}{\partial J^2}Z(J)|_{J=0},\quad
S(J)=S_{YM}+S_q
+J\sum_{\bf x}(L_{\bf x}+L_{\bf x}^\dag)
\ee
Again we have to perform a double character expansion. The leading graphs are neighbouring 
Polyakov loops tiled with plaquettes and their decorations. This leads to a double series for 
$\chi_L(u,h)$ \cite{sc2}. At the critical parameter values 
$u_c, h_c$, the susceptibility will diverge
with a critical exponent
\be
\chi_L(t)\sim\frac{1}{(t_c-t)^{\lambda}},
\ee
with the parametrisation $u=n\cdot t,
h=\frac{1}{n}\cdot t$. We now model the series for the Dlog of $\chi_L$ by Pad\'e approximants in $t$,
which have simple poles at the critical parameter values with the exponent $\lambda$ as residue.
We find $\lambda=1.03$, which signals 3d Ising universality with 
$\gamma=1.237$.
Fixing to the exact value of the exponent, 
the corresponding critical parameters and masses
can be extracted more accurately. Note that 
$\kappa_c\sim O(10^{-2})$ and hence leading order in the hopping
parameter is good. The results for the critical masses for a coarse $N_t=1$ lattice are shown
in \fig\ref{fig_hc_mu}. In particular it is possible to compute the change of the critical mass with 
real and imaginary chemical potential. 
This establishes that in QCD with heavy quarks the deconfinement transition weakens with 
real chemical potential, in full accord with numerical 
findings from the Potts model \cite{potts}.
%


%

\end{document}